\documentclass[10pt]{article}
\pdfoutput=1

\usepackage{graphicx}
\usepackage{subfigure} 
 \usepackage{epsfig}
\graphicspath{{figs/}}
\usepackage{float}

\usepackage[T1]{fontenc}
\usepackage{amsmath, amsthm, amssymb}
\usepackage[ansinew]{inputenc}
\usepackage{amsmath}
\usepackage{xcolor}
\usepackage{soul}

\usepackage{cite}

\usepackage{color}

\usepackage{setspace} 

\usepackage[labelfont=bf,labelsep=period,justification=raggedright]{caption}

\makeatletter
\renewcommand{\@biblabel}[1]{\quad#1.}
\makeatother

\date{}

\begin{document}

\begin{flushleft}
{\Large
\textbf{Inferring models of bacterial dynamics toward point sources}
}
\\
H. Jashnsaz$^{1}$, 
T. Nguyen $^{2}$,
H. I. Petrache $^{1}$, 
S. Press\'e$^{1, 3,\ast}$
\\
\bf{1} Physics Dept., Indiana Univ. - Purdue Univ. Indianapolis, Indianapolis, IN, 46202, US
\\
\bf{2} Stark Neuroscience Institute, Indiana Univ. School of Medicine, Indianapolis, IN 46202, US
\\
\bf{3} Dept. of Cell and Integrative Physiology, Indiana Univ. School of Medicine, Indianapolis, IN 46202, US
\\
$\ast$ E-mail: stevenpresse@gmail.com
\end{flushleft}

\section*{Abstract}
Experiments have shown that bacteria can be sensitive to small variations in chemoattractant (CA) concentrations. 
Motivated by these findings, our focus here is on a regime rarely studied in experiments:  
bacteria tracking point CA sources (such as food patches or even prey). In tracking point sources,
the CA detected by bacteria may show very large spatiotemporal fluctuations which vary with distance from the source.
We present a general statistical model to describe how bacteria locate point sources of food 
on the basis of stochastic event detection, rather than CA gradient information. 
We show how all model parameters can be directly inferred from single cell tracking 
data even in the limit of high detection noise.
Once parameterized, our model recapitulates bacterial behavior around point sources such as the ``volcano effect".
In addition, while the search by bacteria for point sources such as prey may appear random, 
our model identifies key statistical signatures of a targeted search for a point source 
given any arbitrary source configuration.


\section*{Author Summary}
We present a theoretical framework to model bacteria as they track and move toward point chemoattractant (CA) sources (such as small patches of food or prey).
Unlike artificially created gradients, in this regime bacteria must locate the point source despite
low CA concentrations and, thus, high CA spatiotemporal fluctuations.
Using maximum likelihood techniques, we demonstrate that we can infer all model parameters 
directly from single cell tracking data even in the limit of high detection (or external) noise. 
That is to say, even if CA detection events by the bacterium are rare.
Beyond recapitulating known properties of bacteria tracking point sources (such as the ``volcano effect"), 
we use our model to predict statistical signatures of targeted search by bacteria
in more complex environments. Our `top-down' modeling approach is applicable across a broad range of bacterial species.

\section*{Introduction}

Bacteria sense chemoattractants (CA) or chemorepellents (CR) through a sequence of stochastic detection events at their chemoreceptors \cite{Shahrezaei2008369, yuhai2010} and convert temporal variations in the number of detection events into a directional bias \cite{temcomp, bergnature, bergmotion}. 

Experiments report a sensitivity in {\it E. coli}'s response to CAs
down to a few detection events \cite{sourjik, systemschemotaxis}. 
For instance, bacterial runs in {\it E. coli} can be substantially lengthened (by $30\%$)
even in $nM$ gradients \cite{sourjik, systemschemotaxis}.

This suggests that the external noise in the stochastic detection process  -- the `hit' events at the bacterium's chemoreceptors -- 
may affect a bacterium's search strategy for food. 

Here, we are motivated by this work to tackle a regime rarely studied in the literature \cite{2015}: 
how bacteria detect and move toward point food sources -- such as patches of CAs \cite{patches} or even prey or lysed cells \cite{predatorprey} --
where the fluctuations in the number of hits (i.e. external noise) may be very high especially far from the source. 
Beyond high fluctuations in CA concentration away from the source, 
the mean CA concentration emitted from the point source varies very rapidly near the source. 
What is more, point sources -- which generate non-monotonic CA/CR concentration profiles --
can be dynamical (if sources are moving bacterial prey) and be present in large numbers.
These defining characteristics of the CA profile [high fluctuations away from the source, rapidly varying mean near the source]
-- different
from the well-defined CA/CR gradient  \cite{bergnature, yuhai2010, axel, sourjik2} -- give rise to unique bacterial dynamical behavior near the point source.

Our goal is to build a `top-down' model valid across bacterial species that will  
describe how bacteria respond to stochastic detection events (hits) to locate point sources.
One of the main goals of our model will be to identify -- from the dynamics of bacteria near the unique profile setup by point sources -- statistical signatures of targeted search by bacteria toward (or away) from point sources. 
This will help distinguish a random search strategy -- as, for example, is believed to be the case for the 
hunting strategy of the model bacterial predator {\it Bdellovibrio bacteriovorus} --
from an otherwise targeted search for prey \cite{straley_chemotaxis_1977}.

By contrast to the point source regime, much of what is known about chemotaxis is derived from studies on
{\it E. coli} \cite{bergmotion, bergswim, systemschemotaxis, bergimage, temcomp, transienceberg, Eisenbach, Adler21061974, systemsbiology, runtumbletimes, axel, flagella} and often in well controlled, ${\mu}M$, CA gradients \cite{logsensing, bergnature, yuhai2010, noisefiltering}.
For instance, it is known that {\it E. coli} shows an approximate two-state {\it run-and-tumble} dynamics \cite{bergnature, bergswim, yuhai2013, noisefiltering, axel}
generated by the intermittent coalescence and unbundling of its flagella 
which, in turn, is induced by the rotational bias of motors located at each flagellum's base \cite{bergnature, bergswim, bergimage, axel}. 
This simplified model may be nuanced by the stochastic reality that not all motors rotate in lockstep \cite{bergimage}.


As opposed to other modeling approaches \cite{masson}, 
our model will not assume a two-state ({\it run-and-tumble}) dynamics from the onset.
Rather, our model will be constructed starting from simple general principles:  
{\it i)} adaptation (which is the sensitivity to relative rather than absolute changes in CA/CR known to hold in {\it E. coli} \cite{yuhai2010, yuhai2013, logsensing}) and {\it ii)} stochastic signal integration over time through a memory (alternatively `response') function 
entirely determinable from the data  \cite{memory-berg, transienceberg}. 

One key strength of our approach will be 
to show that -- even in the limit of large noise -- all model
parameters can be directly inferred from single cell tracking data using a maximum likelihood approach. 

Once parametrized using one food source configuration (even if it is an artificially well-controlled source),
we will show that the parametrized model is transferable to different and even poorly controlled food source configurations
and can be used to make predictions about dynamical behavior near any source configuration.


\section*{Materials and Methods}
\subsection*{The Model}
{\bf Modeling a point source }\\
We consider a point food source, located at ${\bf r}_{s}$, from which particles are emitted with rate $\mathcal{R}$. 
The particles diffuse away from the source according to the following normal diffusion equation  \cite{infotaxis}
\begin{equation}
\frac{\partial c({\bf r}_{j}|{\bf r}_{s}; t)}{\partial t} = D\bigtriangleup c({\bf r}_{j}|{\bf r}_{s}; t) -  \frac{1}{\tau}c({\bf r}_{j}|{\bf r}_{s}; t) 
+ \mathcal{R}\delta({\bf r}_{j}-{\bf r}_{s})
\label{eq:diff}
\end{equation}
where $\tau$ is the particle decay time constant (which, on physical grounds, can be very large), $D$ is the particle diffusion coefficient
and $\bigtriangleup$ is the Laplacian.
In the most general case, the location of the source is a function of time, ${\bf r}_{s}= {\bf r}_{s}(t)$. The detection rate (called hit rate), $R({\bf r}_{j}|{\bf r}_{s};t)$, 
by the searcher of those particles (the bacterium) is obtained from $c(r_{j}|r_{s};t)$ \cite{infotaxis}.

Here we illustrate the explicit form for $R({\bf r}_{j}|{\bf r}_{s};t)$ for a stationary  concentration profile with open boundary conditions \cite{infotaxis}. In 3 dimensions, we have
\begin{equation}
R({\bf r}_{j}|{\bf r}_{s})= 4\pi aD c({\bf r}_{j}|{\bf r}_{s})=\frac{a\mathcal{R}}{|{\bf r}_{j}-{\bf r}_{s}|}\exp\left(-\frac{|{\bf r}_{j}-{\bf r}_{s}|}{\lambda}\right)
\label{eq:relate}
\end{equation}
where ${\bf r}_{j}$ is the location of the searcher, $a$ is the searcher's radius
and $\lambda = \sqrt{D \tau}$.

In general, the number of hits, $h_{j}$, detected by the searcher at position ${\bf r}_{j}$ over some time interval $[t,t+\Delta T]$ 
is Poisson distributed
\begin{equation}
P(h_{j})=\frac{\left(\int_{t}^{t+\Delta T}dt' R({\bf r}_{j}|{\bf r}_{s};t')\right)^{h_{j}}}{h_{j}!}\exp\left( -\int_{t}^{t+\Delta T}dt' R({\bf r}_{j}|{\bf r}_{s};t') \right).
\label{eq:prv}
\end{equation}
For a fixed source, the above simplifies to
\begin{equation}
P(h_{j})=\frac{\left(\Delta T R({\bf r}_{j}|{\bf r}_{s})\right)^{h_{j}}}{h_{j}!}\exp\left( -\Delta T R({\bf r}_{j}|{\bf r}_{s}) \right).
\end{equation}

\hspace{-0.2in}{\bf Modeling the bacterium}\\
Bacteria do not measure gradients directly. Rather, they
detect stochastic hits at their chemoreceptors and use this hit information
to bias their random walk \cite{bergchemorecep, yuhai2013}.

For this reason, we define a transition probability, $p({\bf r}_{j+1} | \{{\bf r}_{i}, h_{i}\}_{i\leq j} )$, for a bacterium to move to a new position
${\bf r}_{j+1} $ which occurs at every time step. This transition probability is conditioned on 
the bacterium's previous hit history (which is supplied by the conjugate pairs of variables $\{{\bf r}_{i}, h_{i}\}_{i\leq j}$).

Our transition probability, $p({\bf r}_{j+1} | \{{\bf r}_{i}, h_{i}\}_{i\leq j} )$, is a general mathematical object that is not specific to any bacterial species. 
To help make the form for $p({\bf r}_{j+1} | \{{\bf r}_{i}, h_{i}\}_{i\leq j} )$ concrete, we draw from the following physical observations:\\
{\bf 1)} Bacteria show adaptation \cite{signalingmotor, excitationmodelchemotaxis, yuhai2010, systemsbiology} 
(that is, they respond to relative changes in hits not absolute changes) 
and compare hits at different locations to bias their search \cite{bergnature}. 
Thus, their new position, ${\bf r}_{j+1}$, depends on $\bigtriangledown \log h$ not simply $\log h$ or $h$.  We define $\bigtriangledown \log h$
for a discrete $h$ further below.\\
{\bf 2)} Bacteria are subject to random, Brownian, motion \cite{bergrandomwalk, noisepatnaik} as well as internal noise
originating from the stochasticity in relaying their chemotactic signal (such as, for example in {\it E.coli}, binding of active CheY-P to the flagellar motor  
complex which, in turn, biases the motor's rotational direction) \cite{logsensing, sourjik}.
Therefore bacteria can -- at best -- select their new position and direction
from their current position and their past history to within some precision we call $\sigma$.\\
{\bf 3)} Bacteria incorporate previous hit information using a memory function, described below, labeled $\{\alpha\}$
which we will extract from the data \cite{memory-berg, transienceberg}.

These considerations motivate the following general form for the transition probability
\begin{equation}
p({\bf r}_{j+1} | \{{\bf r}_{i}, h_{i}\}_{i\leq j})=\mathcal{N} \exp\left( -\frac{\left({\bf r}_{j+1} - {\bf r}_{j} - \sum\limits_{i=0}^{m}\alpha_{i}{\bf f}_{j-i} \right)^{2}}{2\sigma^{2}}\right),
\label{eq:lik}
\end{equation}
where the coefficients $\{\alpha\}$ -- having dimensions of length --
determine precisely how previous hit information biases the cell's most likely future position ${\bf r}_{j+1}$ and 
\begin{equation}
{\bf f}_{j} \equiv \frac{({\bf r}_{j} - {\bf r}_{j-1})}{|{\bf r}_{j} - {\bf r}_{j-1}|}\cdot \frac{({h}_{j} - {h}_{j-1})}{{h}_{j}} \equiv \bigtriangledown \log h_{j}
\label{eq:cj}
\end{equation}
where, as before, $h_{j}$ are the number of hits at position ${\bf r}_{j}$ where, to be clear, the hits are the number of stochastic detections of CA/CR molecules 
by outer membrane chemoreceptors. 
For convenience, we can write $p({\bf r}_{j+1} | \{{\bf r}_{i}, h_{i}\}_{i\leq j})$ as $p({\bf r}_{j+1} | \{{\bf r}_{i}, \bigtriangledown \log h_{i}\}_{i\leq j})$.
The normalization constant is\\ 
$\mathcal{N} = \int d{\bf r}_{j+1} p({\bf r}_{j+1} |\{{\bf r}_{i}, \bigtriangledown \log h_{i} \}_{i\leq j} )$ and, finally,
$m$ (the `memory') determines how far into the past hit information is considered by the bacterium in selecting its future position. 

The vector 
\begin{equation}
\frac{{\bf r}_{j} - {\bf r}_{j-1}}{|{\bf r}_{j} - {\bf r}_{j-1}|}
\end{equation}
introduced in Eq.~(\ref{eq:cj}) determines the direction in which the motion is being biased. 

As a technical aside, we note that if $h_{j}$ is very small -- and, thus, could be zero -- or if 
sampling a future position in discrete space on a lattice (where the probability of sampling  ${\bf r}_{j+1} = {\bf r}_{j}$ is finite)
then Eq.~(\ref{eq:cj}), could be substituted for this expression 
\begin{equation}
{\bf f}_{j}=\frac{({\bf r}_{j} - {\bf r}_{j-1})}{|{\bf r}_{j} - {\bf r}_{j-1}| + a}\cdot \frac{({h}_{j} - {h}_{j-1})}{{h}_{j} + 1} \equiv \bigtriangledown \log h_{j}
\label{eq:cj2}
\end{equation}
However, in all of our calculations below this modification will not be needed. This is because our 
$h_{j}$ has vanishingly small probability of being $0$ within $\Delta T$ (where $\Delta T$ can be the camera's frame rate in a tracking experiment)
and we also sample positions in continuous space (where the probability of sampling ${\bf r}_{j+1} = {\bf r}_{j}$ is, likewise, vanishingly small).  

Now, we show how all model parameters, $\{ \{\alpha \},\sigma \} \equiv \{ \{\alpha_{0},\cdots, \alpha_{m}\},\sigma \}$, can be directly inferred from single cell tracking data.

\vspace{0.1in}
\hspace{-0.2in}{\bf Parameter inference from single cell tracking data}\\
We assume the following are known from microscopy tracking data: 
1) the searcher's location (e.g. labeled bacterium) and 2), if present, the source location(s) (e.g. locations of patches of food).

To parametrize $\{ \{\alpha\},\sigma \}$, we first write the likelihood of observing a particular bacterial trajectory
\begin{equation}
\mathcal{L}(\{\alpha\}, \sigma | \{ {\bf r}_{i}, \bigtriangledown \log h_{i}\}_{i\leq j}) = \prod_{j} p({\bf r}_{j+1} | \{ {\bf r}_{i}, \bigtriangledown \log h_{i}\}_{i\leq j} ).
\label{eq:infer1}
\end{equation}
This likelihood function is parametrized in terms of the
precise number of particles detected (hits) by the searcher at various points along its trajectory. 
While such a quantity is not directly observable, 
the {\it average} number of hits at any given location is known 
because the distance between the source and the searcher is known. 

Thus, while we cannot maximize $\mathcal{L}(\{\alpha\}, \sigma | \{ {\bf r}_{i}, \bigtriangledown \log h_{i}\}_{i\leq j}) $ directly to obtain  $\{ \{\alpha\},\sigma \}$ in practice, we can certainly maximize
\begin{equation}
\begin{aligned}
\langle \mathcal{L}(\{\alpha\}, \sigma | \{ {\bf r}_{i}, \bigtriangledown \log h_{i}\}_{i\leq j}) \rangle
& \simeq \mathcal{L}(\{\alpha\}, \sigma | \{ {\bf r}_{i}, \langle \bigtriangledown \log h_{i} \rangle \}_{i\leq j})\\ 
& \simeq \mathcal{L}(\{\alpha\}, \sigma | \{ {\bf r}_{i}, \bigtriangledown \log \langle h_{i} \rangle \}_{i\leq j})\\
& =\prod_{j} p({\bf r}_{j+1} | \{ {\bf r}_{i}, \bigtriangledown \log \langle h_{i} \rangle \}_{i\leq j} )
\label{eq:infer}
\end{aligned}
\end{equation}
where -- in going from the first to the second equality -- we have made a cumulant expansion and kept the leading order term, and going from second to third equality we used the approximation that $\langle f(h_j) \rangle \simeq f(\langle h_j \rangle)$.
The validity of this approximation will be assessed by first generating synthetic data where $ \{ {\bf r}_{i}, \bigtriangledown \log h_{i}\}_{i\leq j}$ are known exactly
and comparing the parameters $\{ \{\alpha\},\sigma \}$ determined from
the maximization of the exact likelihood function, $\mathcal{L}(\{\alpha\}, \sigma | \{ {\bf r}_{i}, \bigtriangledown \log h_{i}\}_{i\leq j})$, and the approximate  likelihood function,
$\mathcal{L}(\{\alpha\}, \sigma | \{ {\bf r}_{i}, \bigtriangledown \log \langle h_{i} \rangle \}_{i\leq j})$.
As expected, we will find that the approximate likelihood function requires longer time traces before its maximum 
converges to the correct answer.

Now, from the fully parametrized model, we will show in the results section how we can predict the bacterium's dynamical response to
arbitrary food source configurations, food source emission rates and food source dynamics 
(if the food source, say, is a bacterial prey sought by a predatory bacterial searcher).

From our parametrized model, we can also infer statistical distributions that, in some circumstances, would require much more data to fully quantify 
than is necessary to parametrize $\{ \{\alpha \},\sigma \}$. 
These include, just as examples, predictions regarding: 1) the food source `capture radii' 
(the initial searcher-source distance at which the searcher has a 50/50 chance of finding the source in a specific 
number of steps); 2) both tumble angle and run length distributions in the direction of and away from a food source; 
and 3) the bacterium's adaptation time (i.e. how long it takes for the bacterium to respond to $ \bigtriangledown \log h$ 
or, in other words, how many initial  $\alpha$'s  are zero).

For this reason, it is now convenient to introduce working definitions of {\it run-and-tumble} statistics that we will use in the results section. Mathematically, we define these according to a prescription provided by Berg and Brown \cite{bergnature}. 

Bacterial trajectories are random walks made of successive steps where the change in the direction is ${0}^{\circ}$ to ${180}^{\circ}$ from one time step to the next (time steps could be the frame rate of the camera). If multiple successive steps are straight enough, in other words, if the change in direction between multiple successive steps is small enough, they constitute a run. By definition, a run starts when the change in the direction is less than ${35}^{\circ}$ for three successive steps. The end of a run is when the change in direction is more than ${35}^{\circ}$ for two successive turning points, or when it was greater than ${35}^{\circ}$ for one turning point and the average of the two is also greater than ${35}^{\circ}$. In addition, the tumbling angle is defined as the change in the direction from one run to the next.  

\begin{figure}[H]
\centering
\hbox{\hspace{-4ex}\includegraphics[width=170mm]{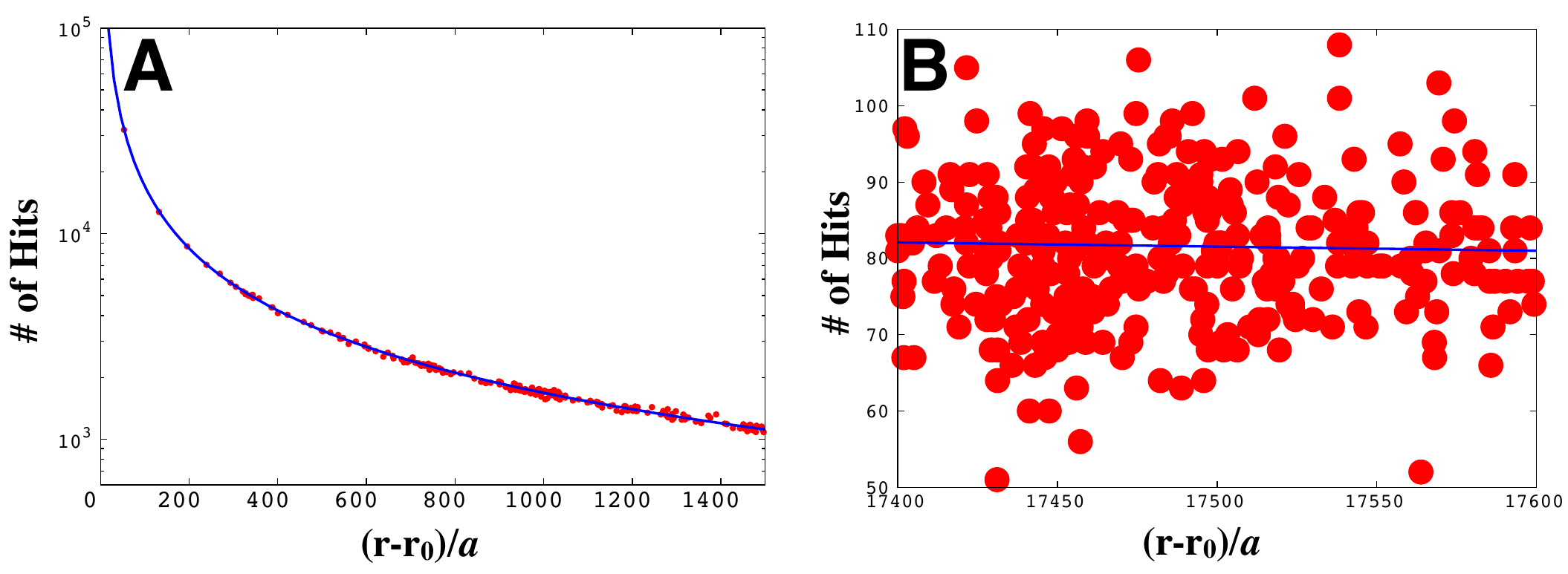}}
\caption{{\bf The noise in the number of hits received by bacteria increases as the distance from the source increases.} 
The red dots in A) denote the exact number of hits the searcher collects over the course of a trajectory moving toward the point source in a log-normal plot.
The blue curve is a plot of 
the mean number of hits expected, 
$\Delta T\times R({\bf r}|{\bf r}_{0})$, plotted against $({\bf r}-{\bf r}_{0})/a$, the radial distance from the source divided by the searcher's radius. 
$R({\bf r}|{\bf r}_{0})$ is given in Eq.~(\ref{eq:relate}). 
In B) we show a region of A) further out from the source [now on a normal plot] exhibiting high fluctuations in the number of hits. 
We used $\Delta T = 0.1s$, $a = 1 \mu m$, $\mathcal{R}=1.7\times{10}^{7} {s}^{-1}$, and $\lambda={10}^{5} \mu m$.}
\label{fig:hits}
\end{figure}

\vspace{0.1in}
\hspace{-0.2in}{\bf Algorithm for generating synthetic data:}\\
To benchmark our method, we generated stochastic bacterial trajectories 
-- that serve as a proxy for single cell tracking data --
following these steps: \\
i) we compute the searcher's mean hits received at its current position ${\bf r}_{j}$ over some interval 
$\Delta T$, $\bar{{h}_{j}} = \int_{t}^{t+\Delta T} dt' R({\bf r}_{j}|{\bf r}_{s};t')$, where $\Delta T$ is an integration time (for example, it can be on the order of $0.1 s$ 
which is a typical tumbling time \cite{axel}).\\
ii) we sample a stochastic hit value, ${h}_{j}$, received at the position of the searcher ${\bf r}_{j}$, from the Poisson distribution ($P(h_{j})=\frac{e^{(-\bar{h}_{j})}(\bar{h_{j}})^{h_{j}}}{(h_{j})!}$). 
As an illustration, the number of stochastic hits plotted versus radial distance from a point source is shown in Fig.~(\ref{fig:hits}); \\
iii) using this hit value (as well as previous hit values and previous positions), we sample the position of the next step, ${\bf r}_{j+1}$, from the transition probability given in Eq.~(\ref{eq:lik}); and \\ 
iv) repeat the previous steps until the searcher reaches the source or, alternatively, a predefined distance from the source. 
Given this synthetic trace, we maximize the likelihood
(or, technically, the log likelihood) with respect to the model parameters $\{\{\alpha\}, \sigma\}$ via a standard grid search 
[by scanning over all possible values
of the parameters and picking those values that maximize the likelihood]. 
We've also maximized our likelihood function using simple Monte Carlo though the real advantage of this approximate method
is realized in cases where we assume a large number of  parameters (i.e. if we have a long memory with many ${\alpha}$'s).
 
\section*{Results}

\begin{figure}[H]
\centering 
\hbox{\hspace{-5ex}\includegraphics[width=170mm]{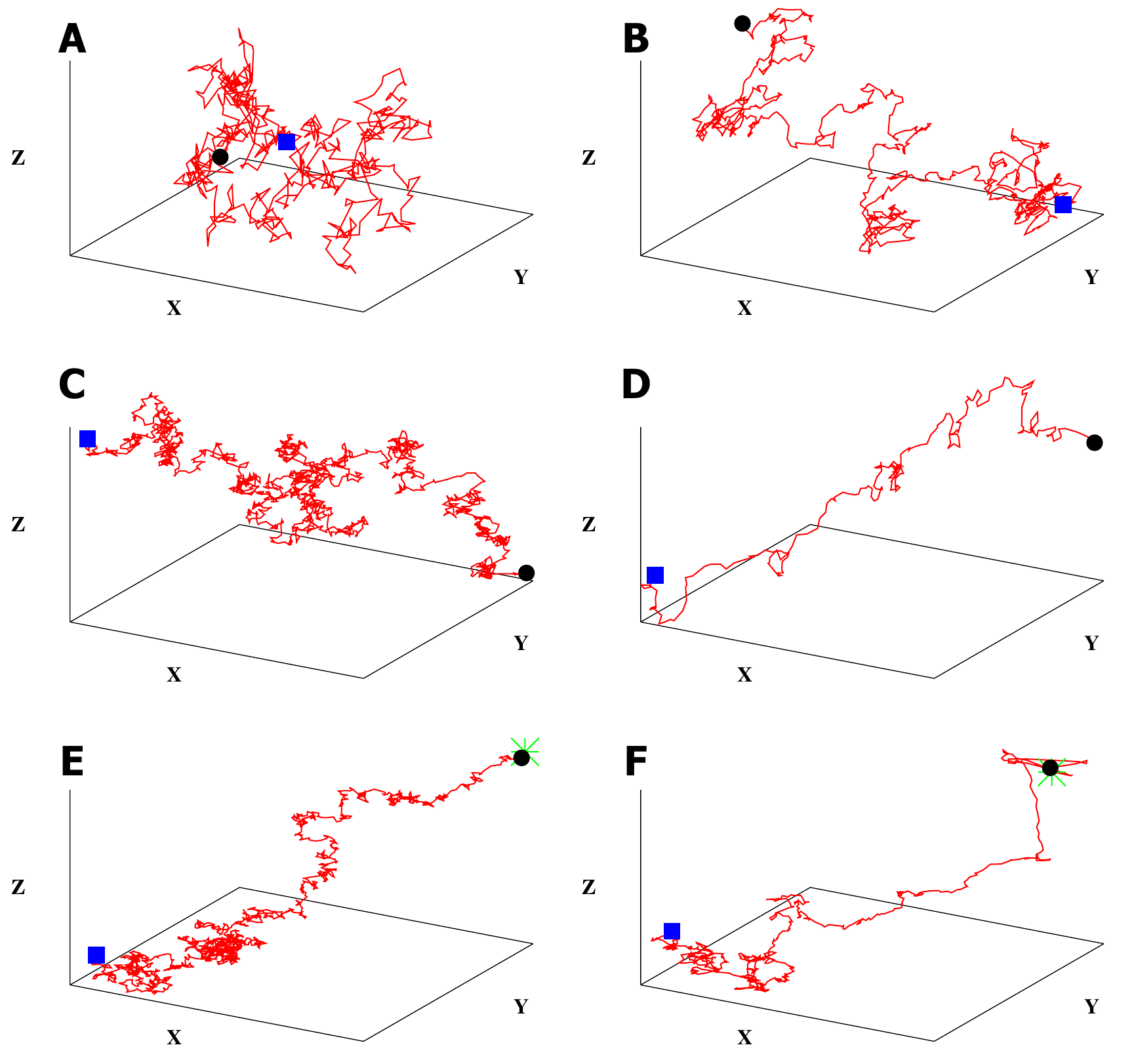}}
\caption{
{\bf Memory of previous hits introduces qualitative differences in bacterial trajectories.} 
All trajectories start from the blue square and end with the black circle.
We have defined $X\equiv x/a$, $Y\equiv y/a$, and $Z\equiv z/a$.
The green star in the last two plots denotes the point source's location. 
Here we show typical bacterial trajectories generated from our model (Eq.~(\ref{eq:lik})) with: 
A) no gradient and one-step memory ; 
B) no gradient and a memory of $m = 5$; 
C) a linear gradient ($x$ direction) and one-step memory ;
D)  a linear gradient ($x$ direction) and memory of $m = 5$;
E) the presence of a food gradient due to a point source at $(1000, 1000, 1000)a$ and one-step memory . 
The searcher starts at $(0,0,0)a$ and locates the source;
F) same as in e) except that we have a memory of $m = 5$. \\
When there is a source, we stop all trajectories when the distance between the searcher and the source 
is less than $60a$. 
We used $\alpha_{0} /a = 130$ 
and $\sigma /a = 10$ throughout.
}
\label{fig:trajectory}
\end{figure}

\begin{figure}[H]
\centering 
\hbox{\hspace{-5ex}\includegraphics[width=170mm]{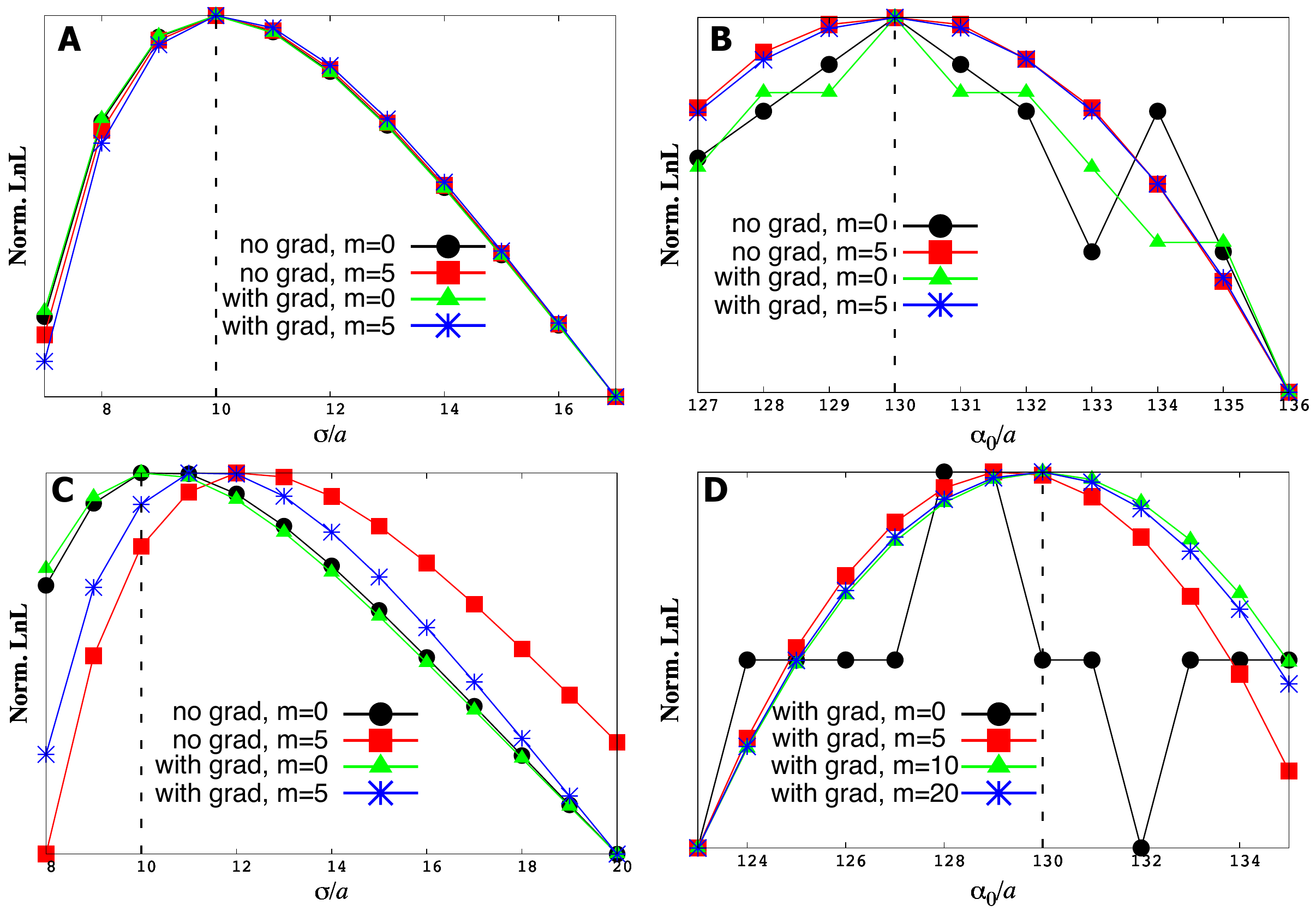}}
\caption{
{\bf We maximize the likelihood function, Eq.~(\ref{eq:infer1}), to find both $\sigma$ and $\alpha_{0}$.} 
As an illustration, we show slices of the likelihood function where one variable is held fixed. For instance, in A) and C) $\alpha_{0}$ 
held fixed while in B) and D) $\sigma$ held fixed. 
In general, we do a full two-dimensional scan to determine both $\sigma$ and $\alpha_{0}$ simultaneously.
Our estimates coincide with the correct theoretical value used to generate the original synthetic trajectory (vertical dashed line
at $\sigma/a = 10$ and $\alpha_{0}/a = 130$).
We tested our method under a variety of conditions.
In particular, `with grad' means in the presence of a point source. A) and B) are inferences made 
using the exact number of hits while C) and D) are made using the average number of hits. 
}
\label{fig:backout}
\end{figure}

\begin{figure}[H]
\centering
\hbox{\hspace{-5ex}\includegraphics[width=165mm]{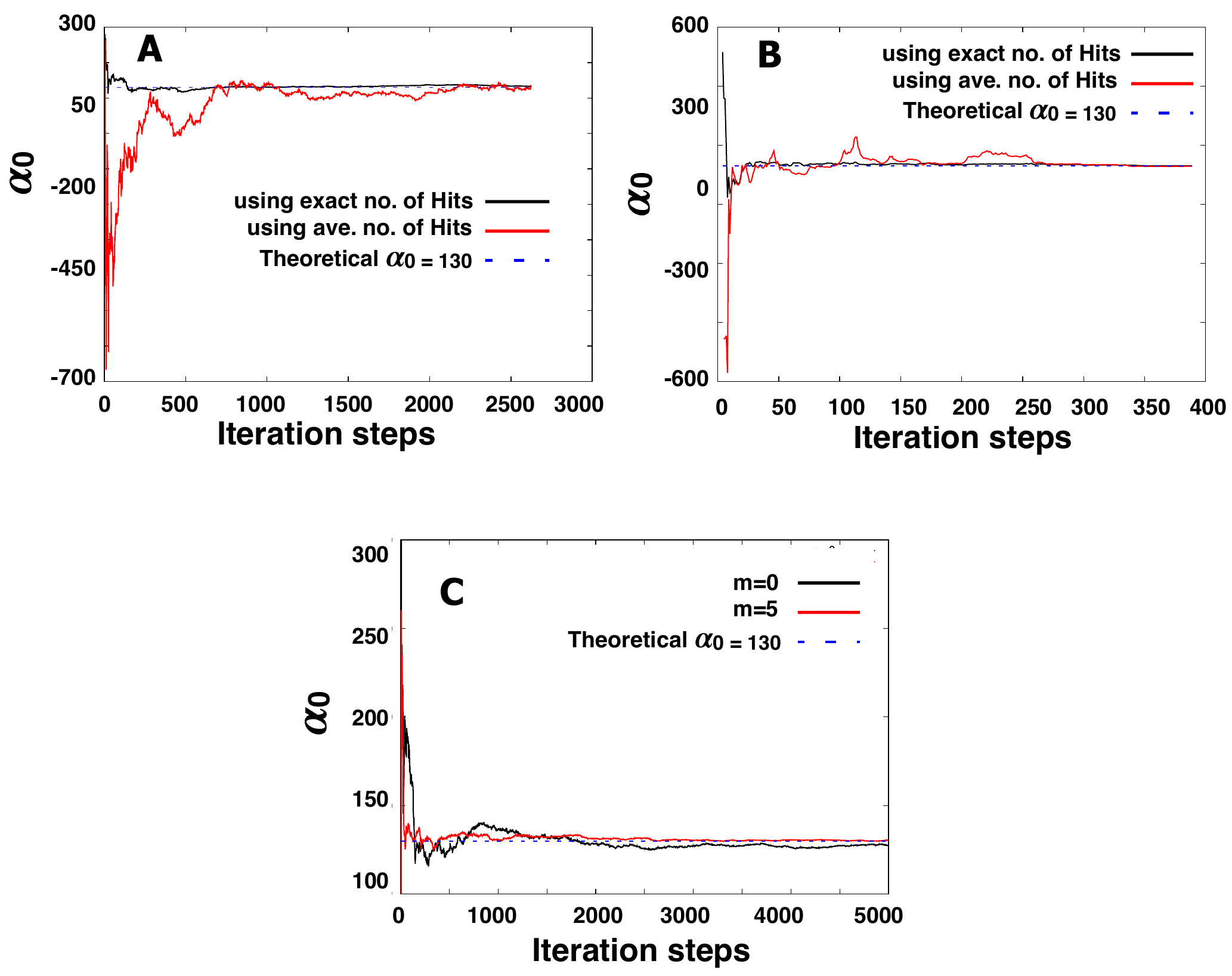}}
\caption{ 
{\bf Within a few hundred iteration steps, our parameter estimates converge to the correct theoretical value
whether we use the exact (Eq.~(\ref{eq:infer1})) or approximate (Eq.~(\ref{eq:infer}))
likelihood function.} Iteration steps are the $\Delta T$'s along the trajectories used to make the point estimate of our parameters.
In A) and B) we consider a point source without memory ($m = 0$) and with memory ($m = 5$), respectively. 
The results of these calculations confirm that our first cumulant approximation of the likelihood function
(shown in Eq.~(\ref{eq:infer})) eventually converges to the correct theoretical parameter values.
In C) we show the same results in the absence of a gradient using the exact number of hits.
}
\label{fig:pointestimate}
\end{figure}

\begin{figure}[H]
\centering
\hbox{\hspace{22ex}\includegraphics[width=75mm]{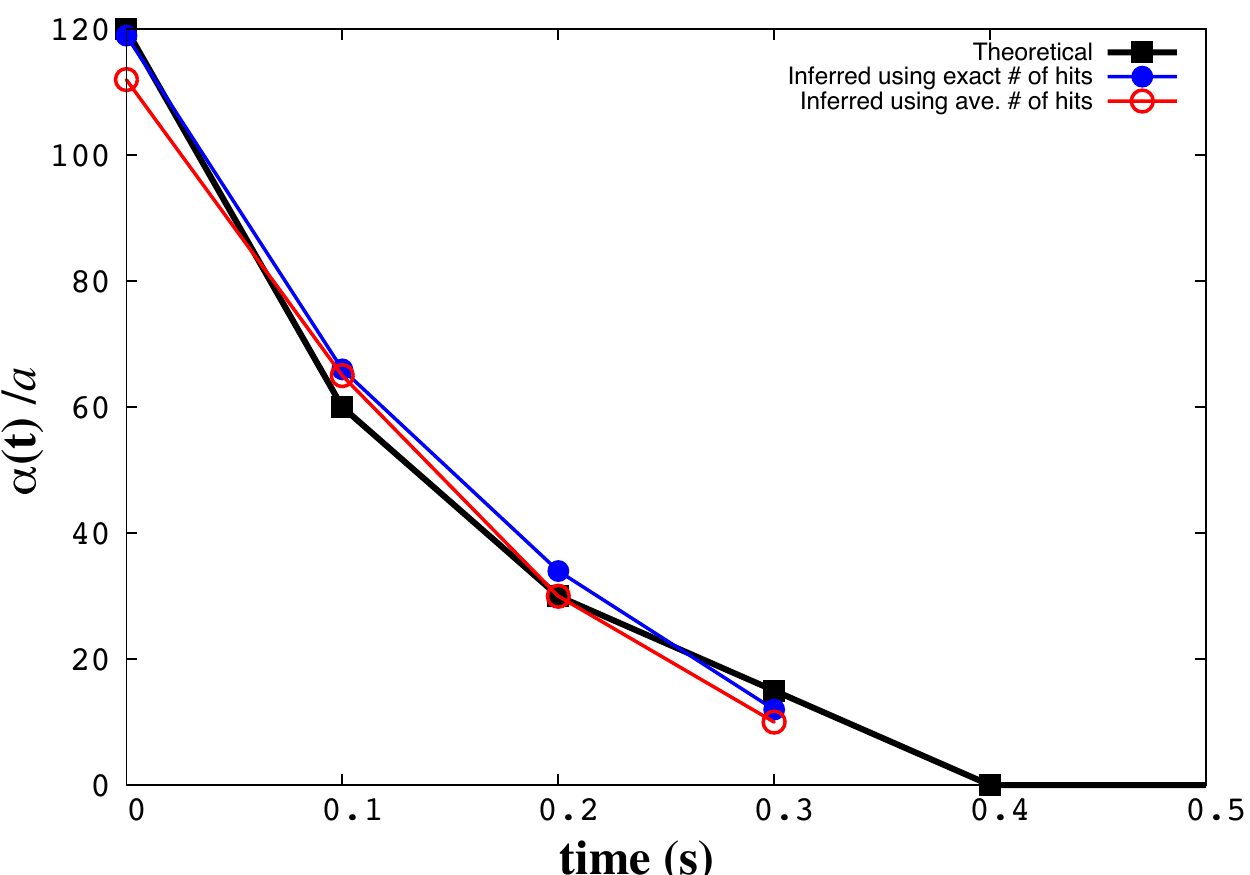}}
\caption{ 
{\bf We can infer model parameters for non-constant memory.} 
We consider the case of decaying memory ($\alpha_i=\frac{1}{2^i}\alpha_0$) and infer ($\alpha_0, \alpha_1, \alpha_2, \alpha_3$) using both exact and average number of hits as indicated in the figure's inset, and compare our estimates to their correct theoretical values used to generate the synthetic data. 
As the number of parameters we need to estimate from the data increases, we need longer trajectories to obtain accurate estimates.
}
\label{fig:backout2}
\end{figure}

\begin{figure}[H]
\centering
\hbox{\hspace{0ex}\includegraphics[width=140mm]{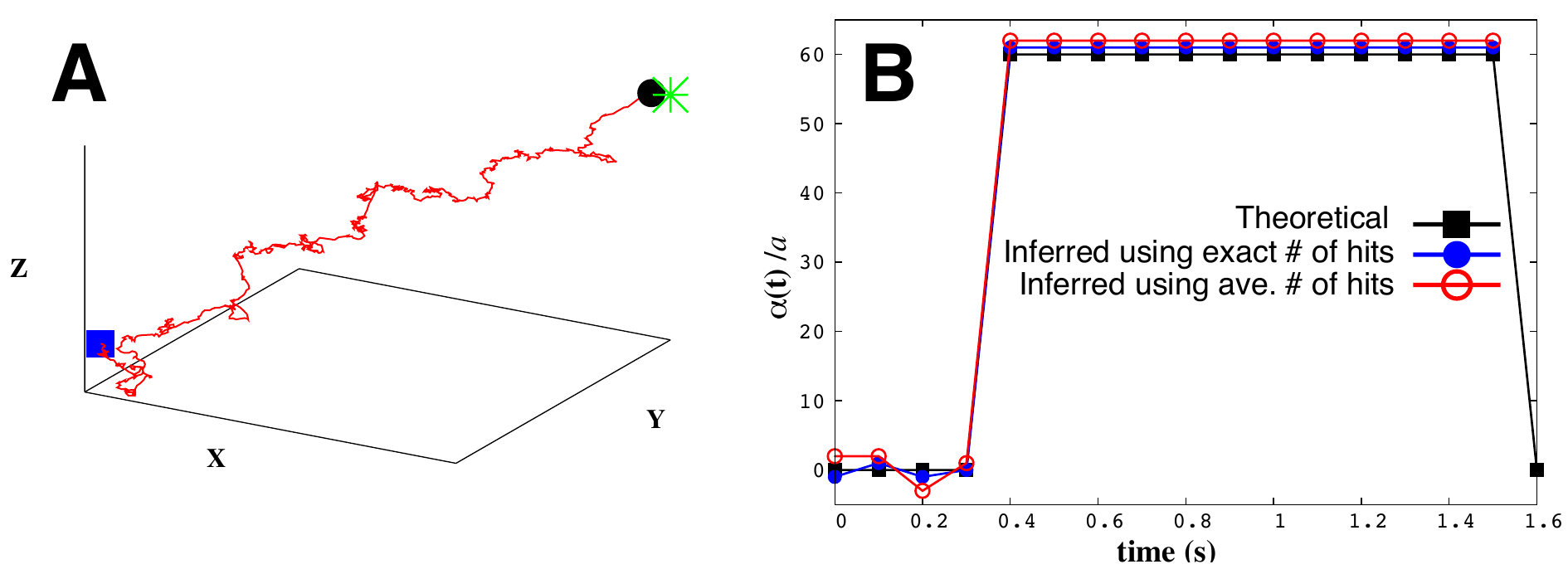}}
\caption{
 {\bf We can infer adaptation times.} 
We consider a theoretical adaptation time 
of $3 \Delta T$ (by setting $\alpha_0 = \alpha_1 = \alpha_2 = \alpha_3 = 0$ and $\alpha_4 = \alpha_5 =  ... = \alpha_{15} = const = 60a$). 
In A) we show a typical trajectory given these $\{\alpha\}$ values. 
In B) we show both theoretical  $\{\alpha\}$'s  (as black squares), our inferred values using the exact hits (as blue dots) as well as 
using the average number of hits (as red circles).
The blue square (in A) 
shows the start of the trajectory at the origin, the black dot shows its end, and the green star denotes the point source's location at $(1000, 1000, 1000)a$. 
We used $\sigma /a = 5$ and stopped the trajectory when the searcher was at a distance $50a$ from the point source. 
We inferred $\alpha_0$ through $ \alpha_3$ individually (as would be necessary in estimating adaptation times from single cell tracking data) 
but assumed $\alpha_4 = \alpha_5 =  ... = \alpha_{15} = const$ and inferred them as a single parameter. 
}
\label{fig:adaptation}
\end{figure}

\begin{figure}[H]
\centering
\hbox{\hspace{-5ex}\includegraphics[width=170mm]{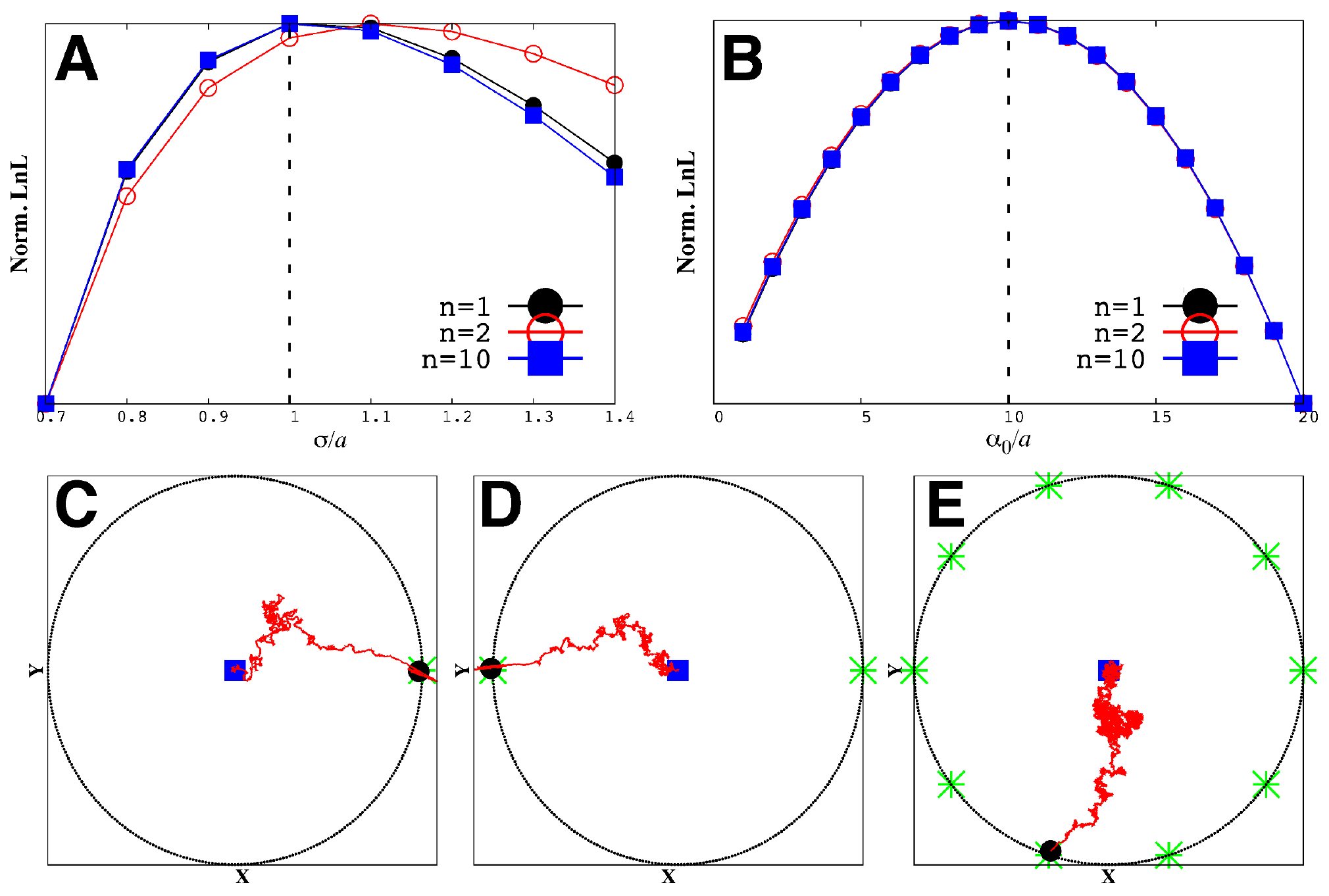}}
\caption{ 
{\bf Our model parameters are insensitive to the particularities of the source configuration.} 
In A and B we show our model parameter inference using the average number of hits for different CA profiles 
($n=1, 2$ and $10$ sources). We infer the same parameters no matter the configuration of the sources around the searcher.
Typical trajectories are shown in C) through E). 
As before, the vertical dashed lines in A and B show the correct theoretical parameters values ($\sigma/a = 1$, $\alpha_{0}/a = 10$). 
We used a memory of $m = 30$ and stopped the trajectory when the searcher was at a distance $5a$ from the point source. Again the blue square shows the start of any trajectory at the origin, the black dot shows its end, and the green star(s) denotes the point source's location. The 
radius, $R$, of the circle on which the point sources -symmetrically- lie in C) through E) is $1000a$. 
The searcher always starts from the center of this circle shown by the blue square.  
}
\label{fig:arbitrary}
\end{figure}

\begin{figure}[H]
\centering 
\hbox{\hspace{-5ex}\includegraphics[width=170mm]{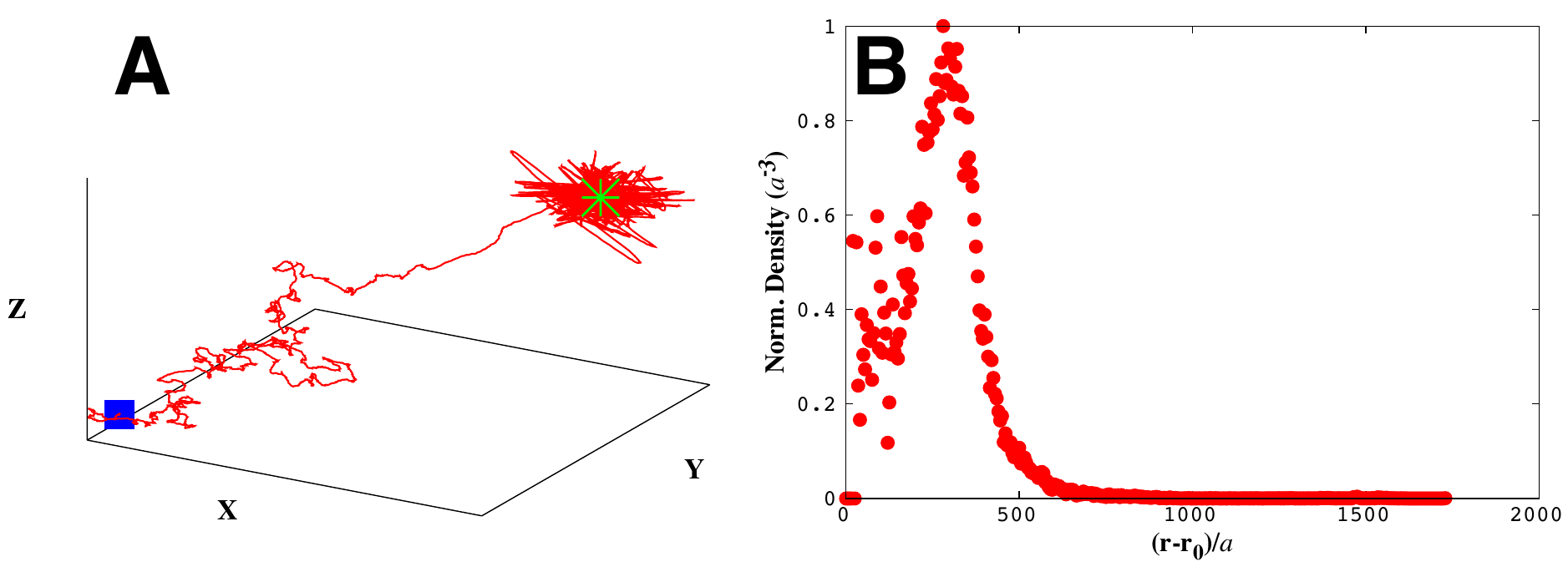}}
\caption{
{\bf  
The volcano effect emerges from our model as a consequence of the rapid CA concentration variation near the source.}
In A) we show a typical trajectory for a bacterium (starting at the origin) exhibiting erratic behavior near the point source ($(1000, 1000, 1000)a$ denoted by the green star). 
In B) we show the probability of finding the searcher as a function of radial distance from the source. The resulting density profile -- resembling the mouth of a volcano in 2d 
\cite{DennisBray1, volcano} -- has maximal bacterial densities occurring on a ring around the point source.
Here we used $m = 30$, $\alpha_{0} /a = 30$, $\sigma /a = 1$ and $X\equiv x/a$, $Y\equiv y/a$, and $Z\equiv z/a$ and stopped the trajectory at
12000 steps. We've normalized the value for the density at $r-r_{0}$ by the volume enclosed in the shell at that distance
to ensure that our results are independent of volume.
}
\label{fig:volcano}
\end{figure}

\begin{figure}[H]
\centering 
\hbox{\hspace{-5ex}\includegraphics[width=170mm]{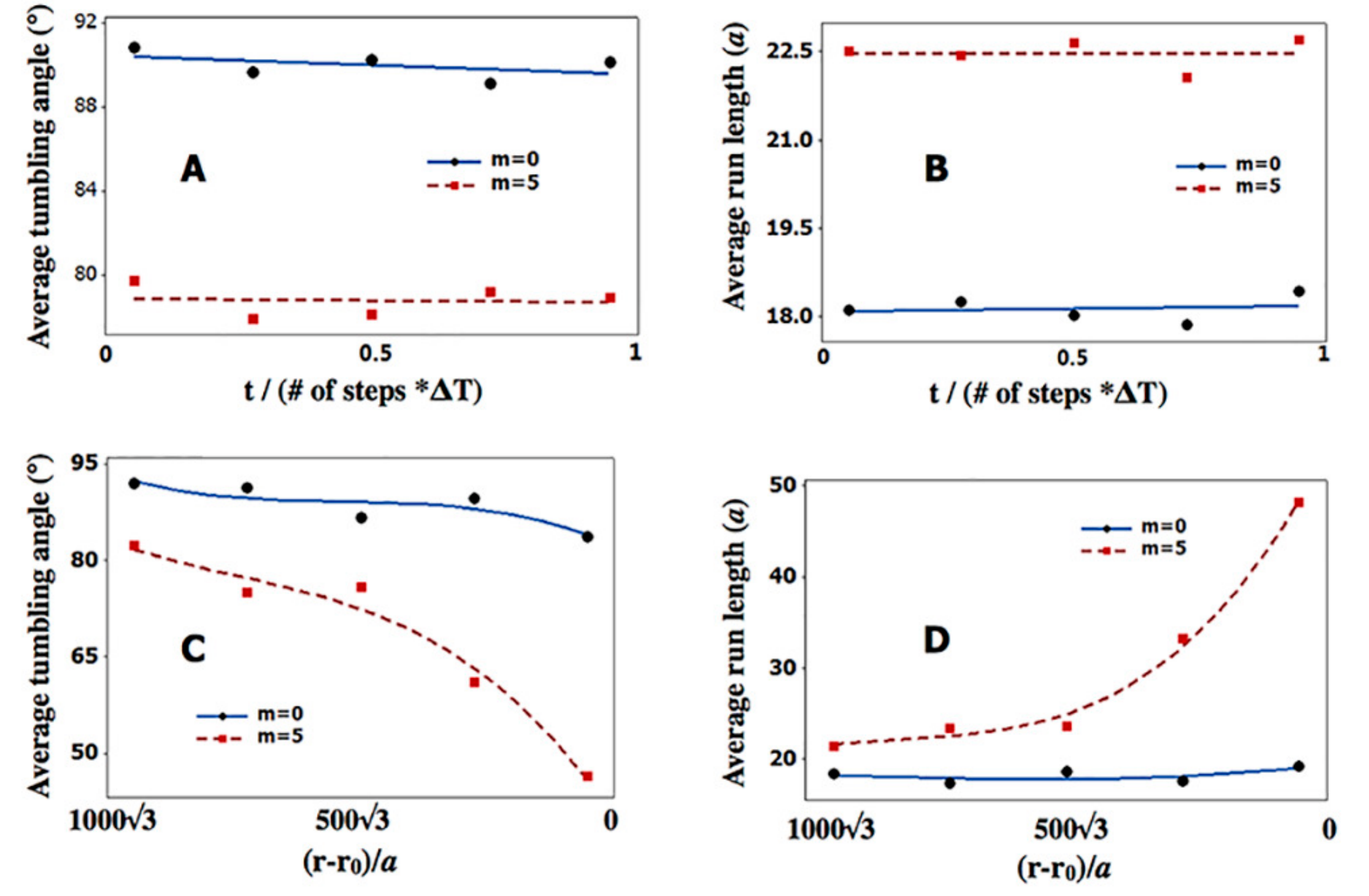}}
\caption{
{\bf {\it Run-and-tumble} statistics are not stationary in the presence of a point food source. They depend on the bacterium's distance from the source.}
The average tumbling angle, in A), and the average run length, in B),
remain roughly constant in time in the absence of gradient. 
We show plots for one step memory ($m=0$) as well as $m=5$.
However, when a point source is introduced, 
the average tumbling angle, in C), as well as the average run length, in D), change as a function of 
distance from the point source
as the searcher moves from the starting point toward the source. 
For all plots we used $\alpha_{0} /a= 130$ and $\sigma /a= 10$ as well as $\Delta T=0.1 s$.
}
\label{fig:averunangle}
\end{figure}

\begin{figure}[H]
\centering
\hbox{\hspace{-5ex}\includegraphics[width=170mm]{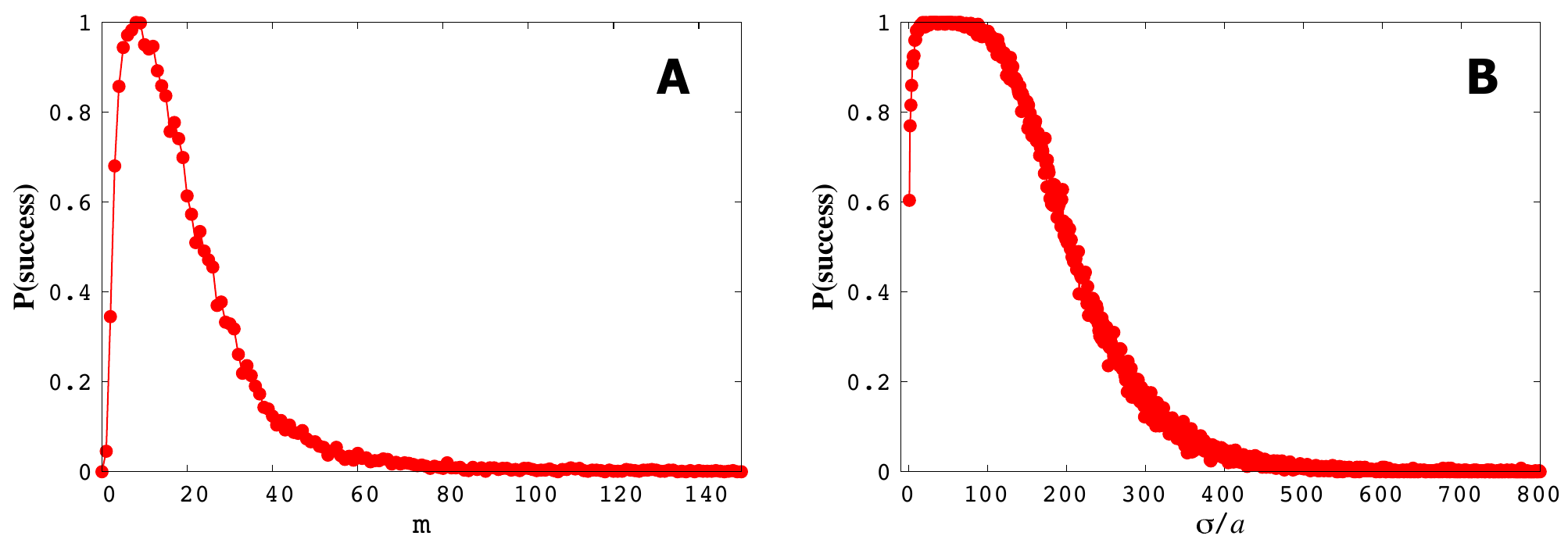}}
\caption{
{\bf The probability of locating a point source is a non-monotonic function of both memory and precision.} 
In A) we show how the probability of successfully locating a source varies with memory, $m$
(using $\sigma /a= 10$ and $\alpha_{0} /a= 130$). 
The searcher starts at $(0, 0, 0)a$ and the point source is located at $(10000, 10000, 10000)a$. 
We call a trajectory successful if, within $50000 \Delta T$ the searcher gets to within $30a$ of the point source. 
In B) we show the probability of success versus $\sigma$ under the same conditions as in A).
We discuss why these distributions are non-monotonic in the main body.
}
\label{fig:memory}
\end{figure}

\begin{figure}[H]
\centering
\hbox{\hspace{-5ex}\includegraphics[width=170mm]{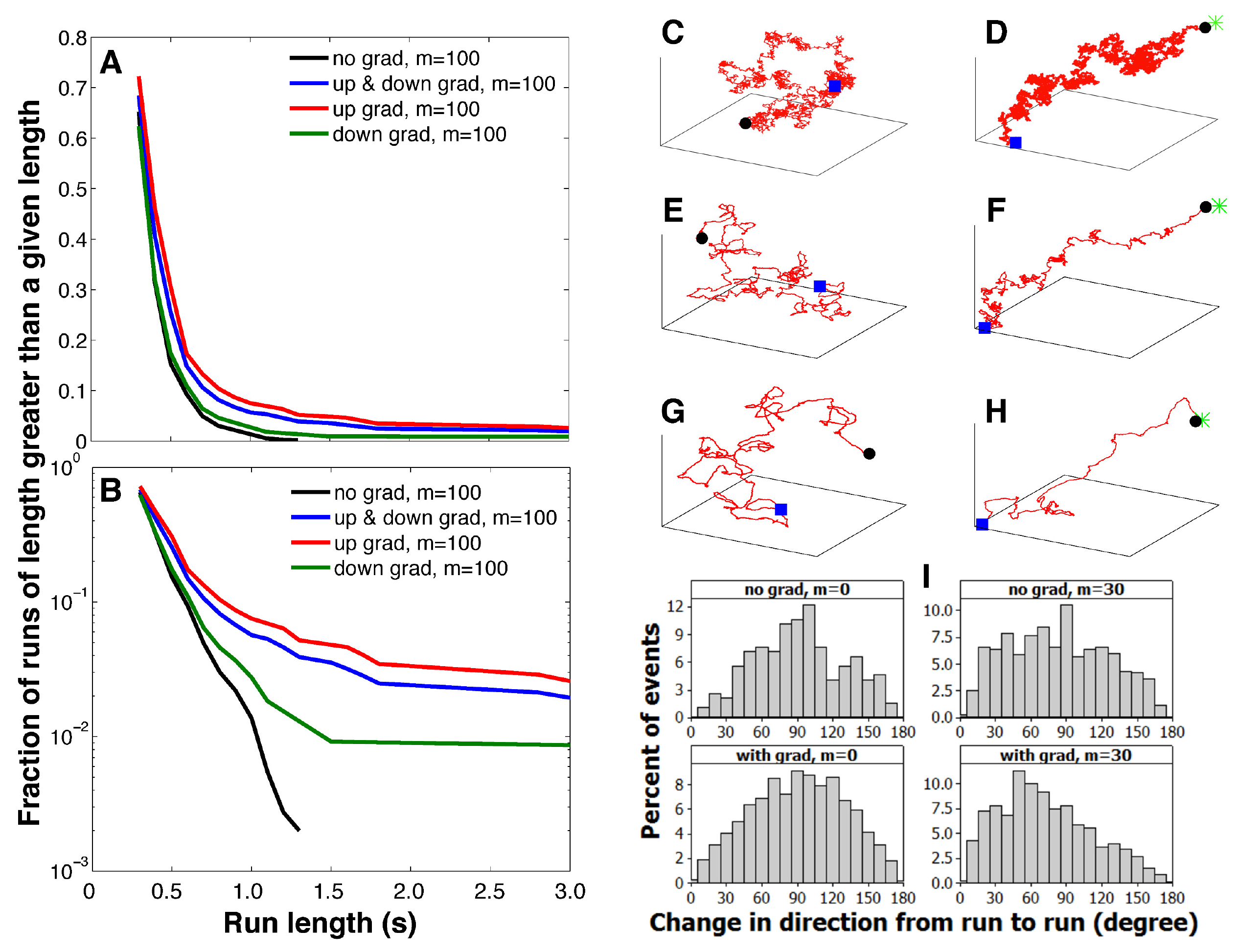}}
\caption{
{\bf Run-and-tumble behavior emerges from our model.}
Here we illustrate how {\it run-and-tumble} statistics emerge from our model by focusing on representative statistics of our trajectories. 
Comparison between the plots described below and experiments are discussed in the main body. 
In A) and B) we show run durations (run lengths in units of seconds) distributions 
in linear and log scale, respectively, 
under various conditions (in the absence and presence of gradient for a memory of $m=100$). 
By construction, the shortest runs are $0.3 s$ long (as per the definition of a run in the main body). 
In C), E), and G) we show the corresponding trajectories in the absence of gradient for the case of one-step memory  ($m = 0$), when $m = 30$, and when $m=100$, respectively. 
In D), F), and H) we show the corresponding trajectories in the presence of gradient due to a point source again for $m = 0$, $m = 30$, and $m = 100$, respectively. 
In I) we histogram changes in direction between the end of one run and the beginning of the next in the absence and presence of gradient with and without memory.
For all trajectories, we used $\alpha_{0} /a = 10$ and $\sigma /a = 1$. 
The point source location is at $(1000,1000,1000)a$. 
Again, the start of all trajectories is shown as a blue square, its end as a black circle. 
The green star denotes the location of the point source. Furthermore, `no grad' indicates that there is no food source and `with grad'
indicates the presence of a point source. Details are discussed in the main body.
}
\label{fig:runangle}
\end{figure}

Our results are broken down along the following topics:\\
1) Role of memory on bacterial behavior, Fig.~(\ref{fig:trajectory});\\
2) Model parameter inference from synthetic single cell tracking data, Figs.~(\ref{fig:backout})-(\ref{fig:adaptation}); \\
3) Predicting bacterial behavior in different source configurations, 
Figs.~(\ref{fig:arbitrary})-(\ref{fig:runangle}).

\vspace{0.1in}
{\bf 1. Role of memory on bacterial behavior:}\\
Before we discuss parameter inference, we briefly highlight
qualitative new trends in Fig.~(\ref{fig:trajectory})  that arise 
in the presence of memory, $m$ as defined in Eq.~(\ref{eq:lik}), that do not explicitly depend on the memory's precise numerical value.
For this reason, in this subsection we only consider bacterial trajectories where the parameters $\alpha_{i}$ are independent of the 
index $i$ and are positive (implying the presence of a CA as opposed to a CR).
 
Figs.~(\ref{fig:trajectory}A)-(\ref{fig:trajectory}B) explore 
the effect of memory from which, as we will show later, emerge {\it run-and-tumble} statistics. 
In particular, Fig.~(\ref{fig:trajectory}A) shows that in the absence of a food gradient with one-step memory -- the case where m=0 from Eq.~(\ref{eq:lik}) -- 
the trajectory is, predictably, a random walk with no preferred direction. However, with memory and no gradient as in Fig.~(\ref{fig:trajectory}B), the searcher
shows increased run lengths.

Figs.~(\ref{fig:trajectory}C)-(\ref{fig:trajectory}D) highlight the searcher's behavior in 
the presence of a {\it linear gradient} with and without memory. As expected, the searcher now exhibits a directional bias
(in the direction of increasing food concentration) however, as we will discuss later,  
tumble angles and run lengths are stationary in time with a linear gradient. In the presence of {\it linear gradient}, the searcher
shows decreased tumbling angles and increased run lengths with memory. 

Tumble angles and run lengths are no longer stationary along the trajectory in the presence of a point food source
denoted by the green star in Figs.~(\ref{fig:trajectory}E)-(\ref{fig:trajectory}F). 
We will demonstrate this quantitatively later in Fig~(\ref{fig:averunangle}). 
Also, the bacterium locates the source exclusively through stochastic CA detection. 
The probability of locating the source depends on parameter values (which we later explore in Fig.~(\ref{fig:memory})).

\vspace{0.1in}
{\bf 2. Model parameter inference:}\\
In Fig.~(\ref{fig:backout}) we show the estimates of the model parameters extracted from trajectories such as those shown in Fig.~(\ref{fig:trajectory}). 
The dotted lines (theoretical values used to generate the data) 
are in excellent agreement with the values inferred from the synthetic data.
This agreement, tested for different parameter values, validates our first cumulant approximation, as detailed by Eq.~(\ref{eq:infer}). 
In addition, Fig.~(\ref{fig:pointestimate}) shows the time (or trajectory length) needed for the results to converge.

Furthermore, we considered the case of non-uniform memory.
First, we chose to make the bacterium's memory decay monotonically (that is, ${\alpha}_{i} = {\alpha}_{0}/2^{i}$), and we inferred model parameters, ($\alpha_0, \alpha_1, \alpha_2, \alpha_3$), for the case of $m=3$ as shown in Fig.~(\ref{fig:backout2}). 

Second, we considered the case where the first few ${\alpha}_{i}$'s are zero. This represents the 
physically relevant effect of a finite adaptation time \cite{UAlon, signalingmotor}. 
That is, the case where the bacterium responds to the gradient at some point in the past though not the immediate past. 
Thus, there is a delay in the bacterium's response to  $\bigtriangledown \log h$. 
Fig.~(\ref{fig:adaptation}) is an important result of our paper.
It shows that we can successfully estimate the bacterium's adaptation time (i.e. estimate the bacterium's delay in response to the local gradient).

\vspace{0.1in}
{\bf 3. Predicting bacterial behavior:}\\
Fig.~(\ref{fig:arbitrary}) also captures a central result of our paper. We show
that the model parameters $(\{{\alpha}\}, \sigma)$ we extract are independent of the source configuration
even in the presence of large external noise and non-uniform gradients.
This outcome is critical in proving that models parametrized in one source configuration can be used to make
predictions about other (perhaps more interesting but less well controlled) source configurations.

Thus, concretely, the information we gather on the model parameters from a single source 
around the bacterium would be sufficient to predict how the bacterium would behave around two and even ten sources; 
see Fig.~(\ref{fig:arbitrary}) for details. 

We emphasize that by contrast to other inference methods for chemotaxis parameters that rely on well-controlled gradients \cite{masson},
our results hold even if, as is the case of a bacterial predator, 
the bacterium is only attracted to point sources (such as prey) where gradients are not well-defined. 
What is more, we do not impose {\it run-and-tumble} dynamics {\it a priori}.

There are a number of statistical signatures of a targeted search by a bacterium that we can now
quantitate that depend on features such as, for instance, the length scale over which a source's gradient varies dramatically rather
than the particularities of the signaling pathway responsible for chemotaxis.
Here are four such signatures:\\
i) In the immediate vicinity of the source, the searcher's trajectory becomes erratic, see Fig.~(\ref{fig:volcano}). That is, bacteria overshoot the source and turn back.
This arises because the CA gradient varies very rapidly with distance in that neighborhood. For instance, the rate function (Eq.~(\ref{eq:relate})) 
increases by as much as $10\%$ for a small displacement by the searcher of just one body length when it is about ten body lengths away from the source
(see, for example, Fig.~(\ref{fig:hits}A)). 
Interestingly, in 1901, in a capillary tube experiment reminiscent of a point food source, it was shown that bacteria swam past high concentration regions 
neighboring the capillary before turning back \cite{rothert, crosby}. \\
ii) The ``volcano effect", describing how bacteria cluster near but not on a point source \cite{volcano, DennisBray1}, emerges from our model.
As a consequence of the bacterium overshooting the source and re-directing its search, the 
bacterium spends most of its time on the approximate surface of a sphere surrounding the source, see Fig.~(\ref{fig:volcano}). \\
iii) {\it Run-and-tumble} statistics are not stationary as the searcher approaches the point source, see Fig.~(\ref{fig:averunangle}). 
In other words runs, on average, get longer and tumbling angles, on average, get smaller in a predictable way. 
The further away the searcher is from the source, the fewer hits it receives, the more tumbles it takes per unit time interval. 
The change in tumble and run statistics can, just like the volcano effect, be indicative of a targeted search by the bacterium.\\
iv)  Given too large a memory (or too little a $\sigma$), 
the bacterium initially overcommits to a particular direction and requires a prohibitive amount of information to re-direct its search.
Predictably, given too little memory (or, equivalently, too large a $\sigma$) a bacterium searches randomly.
Thus the probability of finding a point source is a non-monotonic function of memory and precision, see Fig.~(\ref{fig:memory}).

We highlight that {\it run-and-tumble} behavior is not imposed on our model by hand. 
Rather this behavior qualitatively emerges as a consequence of memory and either stochasticity of the input
or precision with which the bacterium integrates the input and converts hits into a directional bias. 
In particular we briefly compare our results to the well-established observations in the original {\it run-and-tumble} literature:\\
i) Berg and Brown \cite{bergnature} found that for wild type {\it E. coli}, the distribution of tumble lengths as well as the distribution of run lengths is 
approximately
exponential, the shortest tumbles and the shortest runs being the most probable. Tu and coworkers \cite{yuhai2010} also found similar run length distributions 
(Figs. (4)-(6) in Ref. \cite{bergnature} and Fig. (4B) in Ref. \cite{yuhai2010}). 
These observations are recapitulated in our Figs.~(\ref{fig:runangle}A)-(\ref{fig:runangle}B) which show that, under a broad set of parameter values, the same behavior is also observed from our model. Just as in the real data  (Figs. (6) in Ref. \cite{bergnature}) our run distributions
are not perfectly linear on a log-normal plot. That is, they are not perfect exponentials.\\

ii) Tu and coworkers \cite{yuhai2010} found that overall average run length in an exponential gradient is longer than that in a homogeneous environment.
This is also reproduced in our model as shown in Figs.~(\ref{fig:runangle}A)-(\ref{fig:runangle}B) because runs down a gradient are comparable to runs in homogenous environments while runs up a gradient are typically longer. That is, the overall average run length is increased in the presence of gradients. This is also consistent with 
Berg and Brown's results in Ref. \cite{bergnature} where they observed that up gradient run lengths (i.e. runs that move up the gradient)
are typically longer than the down gradient run lengths while the down gradient run length distribution is similar to that of the run length distribution in the absence of gradient. For example, see Fig. (6) (bottom) in Ref. \cite{bergnature}, and Fig. (4B) in Ref. \cite{yuhai2010}). 
In experiments for some CAs, runs down gradients can be longer than runs without gradient (Fig. (6) (bottom) in Ref. \cite{bergnature}). Nonetheless it still holds that such down gradient runs are still typically smaller than runs up gradient. Our model is consistent with these overall  observations (Fig.~(\ref{fig:runangle}A)-(\ref{fig:runangle}B)).\\
iii) Berg and Brown \cite{bergnature} as well as Buguin and coworkers \cite{axel} studied (experimentally and theoretically, respectively) the distribution of bacterial reorientation during tumbling.  For instance, 
Berg and Brown observed a mean angle change from run to run significantly below $90^{\circ}$
in the presence of a gradient (${62}^{\circ}\pm{26}^{\circ}$ in Fig. (3) in Ref. \cite{bergnature}).
Our tumbling angle distributions in Fig.~(\ref{fig:runangle}) are broadly consistent with these observations \cite{bergnature}
and the breadth of our distribution is sensitive to the precise numerical value assigned to our memory parameters.
By contrast, in the absence of a CA/CR gradients, reorientations of the bacterium are random, resulting in a mean angular change of about ${90}^{\circ}$ between successive steps (runs) and this is also consistent with observation \cite{bergnature}. \\
iv) Finally, by construction, our model is consistent with the observation that effects from various sources of CA or CR are additive \cite{transienceberg}. 

\section*{Discussion}

Berg and Brown's original single particle tracking analysis of {\it E. coli} \cite{bergnature, transienceberg} 
not only shed light on {\it E. coli}'s {\it run-and-tumble} dynamics but also
directly motivated the types of models proposed in subsequent decades
\cite{models, yuhai2010, DennisBray1, DennisBray2, stochmodel, sourjikkollmann, golestanian}. 
Since then, the signaling pathway responsible for {\it E. coli}'s chemotactic response has been extensively studied \cite{models, yuhai2010, excitationmodelchemotaxis, DennisBray1, yuhai-berg} and attention has been focused on internal noise sources arising from the stochasticity of the signaling pathway \cite{Jayajit1, Jayajit2, logsensing}.
 
While the biochemical reactions responsible for chemotaxis in some bacteria are  
well understood \cite{bergnature,yuhai2013, models} 
the chemotactic behavior of others 
-- such as that of the model bacterial predator {\it Bdellovibrio bacteriovorus} that preys upon {\it E. coli} --
remain elusive \cite{BV1, BV2}. 

Here our strategy is to extend the theoretical body of work -- and inference work in chemotaxis in particular \cite{masson} --
to study the regime where external (detection) noise is treated explicitly. 
Since we would like our theory to be valid for bacterial species whose chemotactic signaling network is not well characterized,
we do not treat internal noise sources explicitly. 
Instead, internal noise is treated implicitly through the phenomenological precision parameter ${\sigma}$
which we directly infer from the data. 
In other words, the precision parameter implicitly accounts for the noise along the steps of the complex reaction network responsible for signal transduction
from the chemoreceptors to the bacterium's flagella \cite{noisepatnaik, stochmodel, stochmodel2, 2015}  
as well as the noise due to Brownian motion of the bacterium in its environment. 
As a result, our `top-down' approach should be broadly applicable across bacterial species but cannot make molecular-level predictions. 

As input to our model, we have used the fact that 
bacteria show adaptation \cite{signalingmotor, excitationmodelchemotaxis, yuhai2013, models},
employ a temporal sensing mechanism and have a memory of previous events \cite{temcomp}.
We do not  assume two-state {\it run-and-tumble}
dynamics {\it a priori} either \cite{masson}.

Mathematically, our model captures the bacterium's dynamics using a transition probability, Eq.~(\ref{eq:lik}), 
which selects the bacterium's preferred direction within some precision, ${\sigma}$, given memory coefficients, 
$\{\alpha_{i}\}$, which are all to be determined using an {\it inverse}  (maximum likelihood) approach from single cell tracking data. 
Thus, we avoid indeterminable and unobservable
adjustable parameters that often appear in `forward' modeling methods \cite{models}. 
That is, models whose form or parameters are not explicitly inferred from data.

{\it Run-and-tumble} statistics (including whole distributions over trajectories up and down concentration gradients)
then qualitatively arise from our model from basic, physically motivated, principles of chemotaxis.
What is more, our model captures -- at the whole cell rather than at the biochemical level -- critical features that help establish
statistical signatures of targeted search by bacteria towards point sources (such as motion toward bacterial prey by predatory bacteria if predatory bacteria are attracted to CAs released by the prey).
That is, our model makes explicit predictions about the dynamical behavior of bacteria even if external noise is high.
For instance, the volcano effect emerges from our model as a consequence of the distance over which the gradient varies
neighboring a point source. 
In addition, our model shows that if bacteria are tracking point sources 
then they should show changes in run and tumble statistics as they approach the source that we can theoretically anticipate
from the normal diffusive behavior of the CAs.

More interestingly, our model provides a framework to investigate any arbitrarily complex CA/CR arrangement once 
our model is parametrized.
Thus, it is convenient to parametrize a model in simple (presumably well-controlled) environments
to then make predictions about more complex environments.
In addition, and equally importantly, we can infer adaptation times even in the high external noise limit.

For the moment, our model does not treat source-searcher interaction.
However, it is conceivable that a bacterial prey may detect a bacterial predator and respond.
Our model is, in principle, generalizable to dynamical food sources as well as interacting sources and searchers. 
This direction will be the focus of future work. 

\section*{Acknowledgments}

All authors acknowledge stimulating discussions with Prof. G. G. Anderson.
SP also acknowledges the NSF (MCB 1412259) and a Graduate Student Imaging Research Fellowship from the IUPUI 
Office of the Vice Chancellor for Research. 


\bibliography{chemotaxis2}

\begin{thebibliography}{10}
\providecommand{\url}[1]{\texttt{#1}}
\providecommand{\urlprefix}{URL }
\expandafter\ifx\csname urlstyle\endcsname\relax
  \providecommand{\doi}[1]{doi:\discretionary{}{}{}#1}\else
  \providecommand{\doi}{doi:\discretionary{}{}{}\begingroup
  \urlstyle{rm}\Url}\fi
\providecommand{\bibAnnoteFile}[1]{%
  \IfFileExists{#1}{\begin{quotation}\noindent\textsc{Key:} #1\\
  \textsc{Annotation:}\ \input{#1}\end{quotation}}{}}
\providecommand{\bibAnnote}[2]{%
  \begin{quotation}\noindent\textsc{Key:} #1\\
  \textsc{Annotation:}\ #2\end{quotation}}
\providecommand{\eprint}[2][]{\url{#2}}

\bibitem{Shahrezaei2008369}
Shahrezaei V, Swain PS (2008) The stochastic nature of biochemical networks.
\newblock Current Opinion in Biotechnology 19: 369.
\bibAnnoteFile{Shahrezaei2008369}

\bibitem{yuhai2010}
Jiang L, Ouyang Q, Tu Y (2010) Quantitative modeling of {\it {Escherichia}
  coli} chemotactic motion in environments varying in space and time.
\newblock PLoS Comput Biol 6: e1000735.
\bibAnnoteFile{yuhai2010}

\bibitem{temcomp}
Segall JE, Block SM, Berg HC (1986) Temporal comparisons in bacterial
  chemotaxis.
\newblock Proc Natl Acad Sc 83: 8987.
\bibAnnoteFile{temcomp}

\bibitem{bergnature}
Berg HC, Brown DA (1972) Chemotaxis in {\it {Escherichia} coli} analysed by
  three-dimensional tracking.
\newblock Nature 239: 500.
\bibAnnoteFile{bergnature}

\bibitem{bergmotion}
Berg HC (2004) {\it E. coli} in motion.
\newblock Springer.
\bibAnnoteFile{bergmotion}

\bibitem{sourjik}
Sourjik V, Berg HC (2010) Receptor sensitivity in bacterial chemotaxis.
\newblock PNAS 99: 123.
\bibAnnoteFile{sourjik}

\bibitem{systemschemotaxis}
Sourjik V (2004) Receptor clustering and signal processing in {\it {E.} coli}
  chemotaxis.
\newblock T Microbiol 12: 569.
\bibAnnoteFile{systemschemotaxis}

\bibitem{2015}
Lalanne JB, Fran{\c c}ois P (2015) Chemodetection in fluctuating environments:
  Receptor coupling, buffering, and antagonism.
\newblock PNAS 112: 1898.
\bibAnnoteFile{2015}

\bibitem{patches}
Blackburn N, Fenchel T, Mitchell J (1998) Microscale nutrient patches in
  planktonic habitats shown by chemotactic bacteria.
\newblock Science 282: 2254.
\bibAnnoteFile{patches}

\bibitem{predatorprey}
Balagadde FK, Song H, Ozaki J, Collins CH, Barnet M, et~al. (2008) A synthetic
  {\it {Escherichia} coli} predator-prey ecosystem.
\newblock Mol Syst Biol 4.
\bibAnnoteFile{predatorprey}

\bibitem{axel}
Saragosti J, Silberzan P, Buguin A (2012) Modeling {\it {E. coli}} tumbles by
  rotational diffusion; implications for chemotaxis.
\newblock PLoS ONE 7: 35412.
\bibAnnoteFile{axel}

\bibitem{sourjik2}
Vladimirov N, Lebiedz D, Sourjik V (2010) Predicted auxiliary navigation
  mechanism of peritrichously flagellated chemotactic bacteria.
\newblock PLoS Comp Bio 6: e1000717.
\bibAnnoteFile{sourjik2}

\bibitem{straley_chemotaxis_1977}
Straley SC, Conti SF (1977) Chemotaxis by {Bdellovibrio} bacteriovorus toward
  prey.
\newblock Journal of Bacteriology 132: 628--640.
\bibAnnoteFile{straley_chemotaxis_1977}

\bibitem{bergswim}
Berg HC, Anderson RA (1973) Bacteria swim by rotating their flagellar
  filaments.
\newblock Nature 245: 380.
\bibAnnoteFile{bergswim}

\bibitem{bergimage}
Turner L, Ryu WS, Berg HC (2000) Real-time imaging of fluorescent flagellar
  filaments.
\newblock J Bacteriol 182: 2793.
\bibAnnoteFile{bergimage}

\bibitem{transienceberg}
Berg H, Tedesco P (1975) Transient response to chemotactic stimuli in {\it
  {Escherichia} coli}.
\newblock Proc Natl Acad Sc 72: 3235.
\bibAnnoteFile{transienceberg}

\bibitem{Eisenbach}
Eisenbach M (2004) Chemotaxis.
\newblock Imperial College Press.
\bibAnnoteFile{Eisenbach}

\bibitem{Adler21061974}
Adler J, Tso WW (1974) Decision-making in bacteria: Chemotactic response of
  {\it {Escherichia} coli} to conflicting stimuli.
\newblock Science 184: 1292.
\bibAnnoteFile{Adler21061974}

\bibitem{systemsbiology}
Goryanin II, Goryachev AB (2011) Advances in Systems Biology.
\newblock Springer.
\bibAnnoteFile{systemsbiology}

\bibitem{runtumbletimes}
Alon U, Camarena L, Surette MG, Arcas BA, Liu Y, et~al. (1998) Response
  regulator output in bacterial chemotaxis.
\newblock The EMBO Journal 17: 4238.
\bibAnnoteFile{runtumbletimes}

\bibitem{flagella}
Yonekura K, Yonekura SM, Namba K (1977) Complete atomic model of the bacterial
  flagellar filament by electron cryomicroscopy.
\newblock Nature 424: 643.
\bibAnnoteFile{flagella}

\bibitem{logsensing}
Kalinin YV, Jiang L, Tu Y, Wu M (2009) Logarithmic sensing in {\it
  {Escherichia} coli} bacterial chemotaxis.
\newblock Biophysical Journal 96: 2439.
\bibAnnoteFile{logsensing}

\bibitem{noisefiltering}
Andrews BW, Yi TM, Iglesias PA (2006) Optimal noise filtering in the
  chemotactic response of {\it {Escherichia} coli}.
\newblock PLoS Comput Biol 2: 1407.
\bibAnnoteFile{noisefiltering}

\bibitem{yuhai2013}
Tu Y (2013) Quantitative modeling of bacterial chemotaxis: Signal amplification
  and accurate adaptation.
\newblock Annual Review of Biophysics 42: 337.
\bibAnnoteFile{yuhai2013}

\bibitem{masson}
Masson JB, Voisinne G, Wong-Ng J, Celani A, Vergassola M (2012) Noninvasive
  inference of the molecular chemotactic response using bacterial trajectories.
\newblock PNAS 109: 1802.
\bibAnnoteFile{masson}

\bibitem{memory-berg}
BROWN DA, BERG HC (1974) Temporal stimulation of chemotaxis in {\it {Escherichia} coli}.
\newblock PNAS 71: 1388.
\bibAnnoteFile{memory-berg}

\bibitem{infotaxis}
Vergassola M, Villermaux E, Shraiman BI (2007) {"Infotaxis"} as a strategy for
  searching without gradients.
\newblock Nature 445: 406.
\bibAnnoteFile{infotaxis}

\bibitem{bergchemorecep}
Berg HC, Purcell EM (1977) Physics of chemoreception.
\newblock Biophys J 20: 193.
\bibAnnoteFile{bergchemorecep}

\bibitem{signalingmotor}
Barkai N, Leibler S (1997) Robustness in simple biochemical networks.
\newblock Nature 387: 913.
\bibAnnoteFile{signalingmotor}

\bibitem{excitationmodelchemotaxis}
Spiro P, Parkinson JS, Othmer H (1997) A model of excitation and adaptation in
  bacterial chemotaxis.
\newblock Proc Natl Acad Sc 14: 7263.
\bibAnnoteFile{excitationmodelchemotaxis}

\bibitem{bergrandomwalk}
Berg HC (1993) Random Walks in Biology.
\newblock Princeton University Press.
\bibAnnoteFile{bergrandomwalk}

\bibitem{noisepatnaik}
Patnaik PR (2012) Noise in bacterial chemotaxis: Sources, analysis, and
  control.
\newblock Bioscience 62: 1030.
\bibAnnoteFile{noisepatnaik}

\bibitem{UAlon}
U~Alon NB M G~Surette, Leibler S (1999) Robustness in bacterial chemotaxis.
\newblock Nature 397: 168.
\bibAnnoteFile{UAlon}

\bibitem{rothert}
Rothert W (1901) Beobachtungen und betrachtungen über tactische
  reizerscheinungen.
\newblock Flora 88: 371.
\bibAnnoteFile{rothert}

\bibitem{crosby}
Jennings HS, Crosby JH (1901) Studies on reactions to stimuli in unicellular
  organisms. vii. the manner in which bacteria react to stimuli, especially to
  chemical stimuli.
\newblock Amer J Physiol 6: 31.
\bibAnnoteFile{crosby}

\bibitem{volcano}
Simons JE, Milewski PA (2010) The volcano effect in bacterial chemotaxis.
\newblock Mathematical and Computer Modeling 53: 1374.
\bibAnnoteFile{volcano}

\bibitem{DennisBray1}
Bray D, Levin MD, Lipkow K (2007) The chemotactic behavior of computer-based
  surrogate bacteria.
\newblock Current Biology 17: 12.
\bibAnnoteFile{DennisBray1}

\bibitem{models}
Tindall MJ, Porter SL, Maini PK, Gaglia G, Armitage JP (2008) Overview of
  mathematical approaches used to model bacterial chemotaxis I: The single
  cell.
\newblock Bulletin of Mathematical Biology 70: 1525.
\bibAnnoteFile{models}

\bibitem{DennisBray2}
Zonia L, Bray D (2009) Swimming patterns and dynamics of simulated {\it
  {Escherichia} coli} bacteria.
\newblock J R Soc Interface 6: 1035.
\bibAnnoteFile{DennisBray2}

\bibitem{stochmodel}
Shimizu T, Aksenov S, Bray D (2003) A spatially extended stochastic model of
  the bacterial chemotaxis signalling pathway.
\newblock J Mol Biol 329: 291.
\bibAnnoteFile{stochmodel}

\bibitem{sourjikkollmann}
Kollmann M, Sourjik V (2009) In silico biology: From simulation to
  understanding.
\newblock Current Biology 17: 132.
\bibAnnoteFile{sourjikkollmann}

\bibitem{golestanian}
Bennett R, Golestanian R (2013) Emergent {\it run-and-tumble} behavior in a
  simple model of {\it {Chlamydomonas}} with intrinsic noise.
\newblock Phys Rev Lett 110: 148102.
\bibAnnoteFile{golestanian}

\bibitem{yuhai-berg}
Tu Y, Shimizu TS, Berg HC (2008) Modeling the chemotactic response of
  {\it Escherichia coli} to time-varying stimuli.
\newblock PNAS 105: 14855.
\bibAnnoteFile{yuhai-berg}

\bibitem{Jayajit1}
Mukherjee S, Seok SC, Vieland VJ, Das J (2013) Cell responses only partially
  shape cell-to-cell variations in protein abundances in {\it Escherichia coli}
  chemotaxis.
\newblock NPAS 110: 18531.
\bibAnnoteFile{Jayajit1}

\bibitem{Jayajit2}
Mukherjee S, Seok SC, Vieland VJ, Da J (2013) Data-driven quantification of the
  robustness and sensitivity of cell signaling networks.
\newblock Phys Biol 10: 066002.
\bibAnnoteFile{Jayajit2}

\bibitem{BV1}
Lambert C, Fenton AK, Hobley L, Sockett RE (2011) Predatory {\it Bdellovibrio}
  bacteria use gliding motility to scout for prey on surfaces.
\newblock JOURNAL OF BACTERIOLOGY 193: 139.
\bibAnnoteFile{BV1}

\bibitem{BV2}
Morehouse KA, Hobley L, Capeness M, Sockett RE (2011) Three motab stator gene
  products in {\it Bdellovibrio} bacteriovorus contribute to motility of a
  single flagellum during predatory and prey-independent growth.
\newblock JOURNAL OF BACTERIOLOGY 193: 932.
\bibAnnoteFile{BV2}

\bibitem{stochmodel2}
Alber M, Chen N, Glimm T, Lushnikov P (2003) Multiscale dynamics of biological
  cells with chemotactic interactions: From a discrete stochastic model to a
  continuous description.
\newblock J Mol Biol 329: 291.
\bibAnnoteFile{stochmodel2}

\end{thebibliography}
\bibliographystyle{plos2009.bst}

\end{document}